\def\etal{{\rm et al. }}

\def\kms{{\rm km s^{-1}}}
\newcommand\aap{{\em A}\&{\em A}}

\newcommand\aj{{\em AJ}}
\newcommand\apj{{\em ApJ}}
\newcommand\apjl{{\em ApJ}}
\newcommand\apjs{{\em ApJS}}

\newcommand\mn{{\em MNRAS}}

\newcommand\nature{{\em Nature}}
\newcommand\pasp{{\em PASP}}

\documentclass[structabstract]{aa}  
\usepackage{graphicx}
\usepackage{txfonts}
\usepackage[sort]{natbib}
\bibpunct{(}{)}{;}{a}{}{,}
\begin{document}

\titlerunning{AGN spiral galaxies in groups: effects of bars}
\authorrunning{Alonso et al.}

\title{AGN spiral galaxies in groups: effects of bars}
\subtitle{}

\author{Sol Alonso\inst{1,2},Georgina Coldwell\inst{1,2}
   		  \and
           Diego G. Lambas\inst{1,2,3}
          }

\institute{Consejo Nacional de Investigaciones Cient\'{\i}ficas y T\'ecnicas, CONICET, Argentina
              \email{solalonsog@gmail.com} \and
             FCEFyN, Universidad Nacional de San Juan, San Juan, Argentina \and
            IATE, OAC, Universidad Nacional de C\'ordoba, Laprida 854, X5000BGR, C\'ordoba, Argentina
}
             
   \date{Received xxx; accepted xxx}

  \abstract
   {}
   {We explore properties of barred active spiral galaxies in groups and clusters selected from the SDSS-DR7, with the aim of assessing the effects of bars on AGN and the role of the high density environment.}
   {We identified barred active galaxies that reside in groups and clusters from SDSS-DR7 group catalog.
With the goal of providing a suitable quantification of the effects of bars, a reliable control sample of unbarred active galaxies in high density environments with similar redshift, magnitude, morphology, and bulge size distributions was also constructed.
}
   {We found that the fraction of barred AGN galaxies in groups and clusters ($\approx$ 38$\%$) is higher than those in the total barred AGN sample ($\approx$ 28$\%$), 
indicating that AGN spiral galaxies in groups are more likely to be barred than those in the field.
We also found that barred AGN galaxies are more concentrated towards the group centers than the other unbarred AGN group members.
In addition, barred AGN host galaxies show an excess of population dominated by red colors, with respect to the control 
sample, suggesting that bars produce an important effect on galaxy colors of AGN hosts.
The groups of AGN galaxies with and without bars show similar virial masses; however, the host groups of the barred AGN exhibit a larger fraction of red colors than the host groups of the corresponding unbarred active galaxies.
Color-magnitude relations of both host groups of AGN differ significantly: the host group colors of barred active galaxies display distributions spreading toward red populations, at the same $(M_r)_{Group}$, with respect to the host groups of the unbarred AGN objects. 

Barred active galaxies show an excess of nuclear activity compared to galaxies without bars.
We found that barred active galaxies located farther from the group-center have stronger $Lum[OIII]$.
In addition, for both AGN samples, nuclear activity increases in bluer host groups however, barred active objects systematically show higher $Lum[OIII]$ values, irrespective of the global group colors.
Our findings suggest that the efficiency of bars to transport material towards the more central regions of the AGN galaxies in high density environments reveals an important dependence on the localization of objects within the group/cluster and on the host group colors. 
}
  

\maketitle
%

\section{Introduction}

Galactic bars are believed to be directly related to the dynamical evolution of their host spiral galaxies.
Several studies of numerical simulation show that bars can efficiently transport gas from the outer regions of galaxies to the central 
kiloparcec scale \citep{wei85,deb98,ath03}.
The gas clouds suffer shocks by interaction with the edges of the bars, producing angular momentum losses and allowing a flow of gas toward the innermost central regions of the barred galaxies \citep{SBF90}. 
Given the high efficiency of gas inflow, the presence of bars plays an important role in enhancing the nuclear star formation
\citep{hec80,haw86,deve87,ars89,huang96,HFS97,MF97,emse01,kna02,jog05,hunt08}. 

In addition, the bar-driven material toward the most inner regions has been considered an efficient mechanism for triggering activity in active galactic nuclei (AGN) \citep{cor03,comb93}.
\cite{SBF89} proposed a model that they called "bars within bars" as a mechanism for fueling AGN galaxies. In this scenario, the external bars transport gas on a scale of a few parsec and the material is accumulated.
It shows a gravitational instability producing
a secondary bar, which allows gas to approach closer 
to a central black hole \citep{mul97,mac97,HFS97}.
Different observational studies, using the Hubble Space Telescope images and the spectroscopic data obtained from ISAAC/VLT and SAURON 
\citep{emse01,malk98,laine02,caro02} have confirmed the presence of secondary bars, by observing nuclear bars inside large-scale external bars.
 
There is still a debate on the observational analysis that shows different results regarding the effects produced by bars on the central nuclear activity.
For instance, \cite{HFS97} found that bars have no significant effect on the AGN activity in a sample of 187 barred spiral galaxies.
\cite{lee12} also used a sample of AGN, selected with the \cite{kew01} classification,
from the Sloan Digital Sky Survey Data Release 7 (SDSS-DR7) volume limited sample of late-type galaxies ($b/a > 0.7$) concluding that bars do not have an important effect on the nuclear activity.
However, \cite{oh12} found evidence that bars enhance central activity in blue galaxies with low black hole masses
from a sample of late-type active galaxies ($b/a > 0.6$) selected from SDSS-DR7 with the \cite{kauff03} criteria.
In the analysis, \cite{lee12} do not include the composite galaxies within the sample of AGN 
host and because these galaxies are in general, younger and less massive \citep{kew06} than the AGN 
sample defined by the \cite{kew01} criteria. So, the different AGN selection criteria could 
affect the results related to the effect of enhanced AGN activity  by bars.

More recently, Alonso et al. (2013, hereafter A13) studied a sample of face-on barred AGN spiral galaxies, in comparison with a suitable control sample of unbarred active galaxies with similar redshift, magnitude, morphology, bulge sizes, and local environment distributions.
They found that barred AGN galaxies systematically show 
a higher fraction of powerful activity with respect to unbarred AGN in the control sample. In addition, the analysis of the accretion rate onto the central black holes shows that barred AGN host galaxies show an excess of objects with high accretion rate values with respect to unbarred active ones.

Galaxy properties (star formation, gas content, morphology, etc.) show a strong dependence on the local environment in which galaxies reside. 
Therefore, to have a better understanding of the formation and evolution of bars, an important point is to analyze, not only the internal structures of spiral galaxies, but also the external influence on their host environment.
Nevertheless, the dependence between the bar fraction and environment is a controversial point since different studies show contradictory results.
Several observational works found that the bar galaxy fraction 
does not depend on the environment \citep{vandenB02,mendez10,martinez11,lee12}.
However, \cite{elme90} found a higher bar fraction in pair systems for a sample of spiral galaxies. 
In addition, \cite{esk00} show that the bar fraction in the Fornax and Virgo clusters is slightly higher than the average value for fields. Moreover, from the analysis of galaxies in the Shapley-Ames Catalog, \cite{vandenB02} found that the fraction of barred galaxies in cluster environments is larger than in lower density regions.
Recently, \cite{skibba12} show that barred galaxies tend to reside in denser environments than their unbarred counterparts for a large sample obtained from the 
Galaxy Zoo 2 project \citep{lintott11}, finding a clear environmental dependence on barred galaxies.

Motivated by these finds, here we analyze the effect of bars on active nuclei galaxies and the role played by high density environments.
We study active galaxies with and without bars within groups and clusters with the aim of assessing whether high density environments play a significant role in modifying the host properties and nuclear activity of barred active galaxies.
For this purpose, using data from the SDSS-DR7, we obtain samples of barred and unbarred AGN residing in groups and clusters of galaxies.
We analyze the spatial distribution, host galaxy properties, and nuclear activity of barred AGN in high density regions, and compare them to those of unbarred active galaxies from a suitable control sample, with the aim of obtaining additional clues about the interplay between the high density environments and barred AGN galaxies.

This paper is structured as follows. Section 2 describes the procedure used to construct the catalog of barred AGN group galaxies from SDSS-DR7, and the control sample selection criteria. 
In Section 3, we study the barred and unbarred AGN relative position in groups/clusters. 
Section 4 explores the bar dependence on colors of host active galaxies with respect to both group central distance and group virial masses and Section 5 analyses environmental properties.
In Section 6, we analyze the black hole activity in barred AGN in high density environments, and in Section 7 we summarize our main conclusions. 
The adopted cosmology through this paper is  
$\Omega = 0.3$, $\Omega_{\lambda} = 0.7$, and $H_0 = 100~ \kms \rm Mpc$.


\section{Selection of barred AGN galaxies in high density environments}

The analysis of this paper is based on the SDSS-DR7 \citep{abaz09} photometric and spectroscopic galaxy catalog; DR7 is the seventh data release corresponding to the survey SDSS-II, and comprises 11663 square degrees of sky imaged in five band widths ($u$, $g$, $r$, $i$, and $z$).
It provides imaging data for 357 million objects, as well as spectroscopy over $\simeq \pi$ steradians in the north Galactic cap and $250$ square degrees in the south Galactic cap.
The SDSS-DR7 main galaxy sample is essentially a magnitude limited spectroscopic sample with $r_{lim}<17.77$ covering a redshift range $0<z<0.25$, with a median redshift of 0.1 \citep{strauss02}.
For this work, several physical properties of galaxies were derived from the published SDSS-DR7 data: 
stellar mass content, indicators of star-bursts, emission-line fluxes, concentration index, etc. \citep{brinch04, blanton05}, and were obtained from the \texttt{fits} files at the MPA/JHU \footnote{http://www.mpa-garching.mpg.de/SDSS/DR7/} and the NYU\footnote{http://sdss.physics.nyu.edu/vagc/} catalogs. 

\begin{figure*}
\includegraphics[width=180mm,height=140mm ]{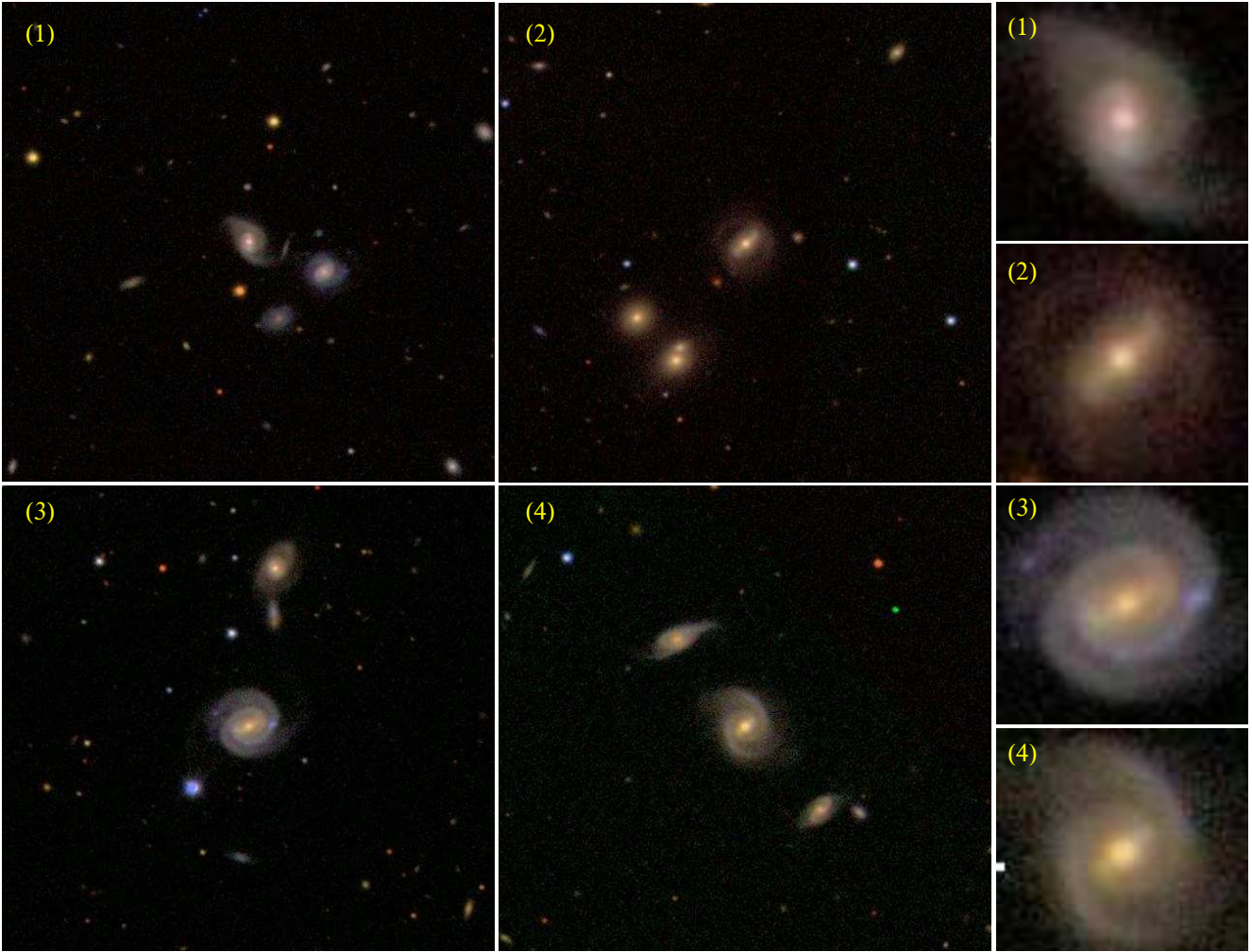}
\caption{Images of typical examples of barred active galaxies in high density environments. 
The images are $\approx$ 3$'$.5 x 3$'$.5. 
Right images show a zoom of the barred AGN galaxies.}
\label{Ej}
\end{figure*}

In our previous work (A13) we compiled a barred AGN catalog obtained from SDSS-DR7. 
The active galaxy sample was selected by applying the \cite{kauff03} criteria, employing a standard diagnostic diagram proposed by \cite{BPT81} (hereafter BPT), and by using the publicly available emission-line fluxes (for details see A13). 
By visual inspection of SDSS images 
we classified 6772 hosts of face-on ($b/a>0.4$) AGN spiral galaxies brighter than $g-$mag$<$16.5 and with redshifts $z<0.1$ as barred or unbarred. 
We found 1927 AGN hosted in barred galaxies, representing a fraction of 28.5$\%$ with respect to the full sample of AGN in spiral host galaxies, and this value agrees with the bar fraction found by visual inspection of optical galaxy samples in previous works 
\citep{deVau91,nilson73,marinova09,master11,oh12}. 
The remaining 4628 face-on spiral galaxies are unbarred. The authors excluded 207 objects that did not match the bar classification completely. 

To analyze the properties of barred AGN galaxies in high density environments in detail, in comparison with AGN without bars, we identified barred and unbarred active objects that reside in groups and clusters by cross-correlating the total barred and unbarred AGN samples with the SDSS-DR7 group catalog constructed by \cite{zap09}.
These authors identified groups of galaxies using a friends-of-friends algorithm with
the same parameters corresponding to the values found by \cite{merch05}
to produce a reasonably complete sample (95$\%$) with low contamination ($<$ 8$\%$).
The group catalog has 122,962 galaxies within 12,630 groups that have a minimum number of four members
 and $z<0.1$.  
 Moreover, it provides several useful galaxy group properties used throughout the paper, such as the geometric group center coordinates, virial mass, richness, line-of-sight velocity dispersion, and virial radius.

As a result of this cross-correlation, we obtained a sample of 319 barred and 519 unbarred 
AGN galaxies in groups. 
These values represent a fraction of 16.6$\%$ with respect to the full sample of 1927 barred AGN galaxies, and 11.2$\%$ with respect to a total of 4628 AGN without bars.
Table 1 lists the numbers and percentages of the different AGN galaxy samples used in this work, and 
here we observed that the fraction of barred AGN galaxies in groups ($\approx$ 38$\%$) is higher than those in the total barred AGN sample ($\approx$ 28$\%$).
This finding shows a higher bar fraction in groups and clusters of galaxies, indicating that AGN spiral galaxies in groups are more 
likely to be barred than those in the field, in agreement with previous work by \cite{skibba12} and \cite{esk00}. 
This suggests that bars are stimulated in accordance with the local density environment and that the physical processes such as ram pressure stripping, starvation or strangulation, harassment and interactions in groups and clusters seem to be related to the bar phenomenon. 
\cite{skibba12} found that bars are formed by secular evolution and that this process depends on the local environment of the host galaxies, suggesting that there is a dichotomy between internal secular mechanisms and external environment.  
Recently, \cite{kor12} show that harassment may influence secular evolution of the bars by the cumulative effect of several rapid encounters with other cluster members.  
In addition, starvation or strangulation, which are gas stripping processes of the hot diffuse 
component of satellite galaxies, influences the star formation on longer timescales \citep{larson80} and could also stimulate the growing bars 
\citep{bere07,master12}.
Figure 1 shows images of typical examples of barred AGN host galaxies in high density environments selected from our sample.

\begin{table}
\center
\caption{Results of classification.}
\begin{tabular}{|c c c| }
\hline
Classification & Number of galaxies &  Percentages  \\
\hline
\hline
Barred             &  1927    &   28.5$\%$     \\
Unclear barred     &  207     &    3.0$\%$     \\
Unbarred           &  4638    &   68.5$\%$     \\
\hline
Total              & 6772     &   100.0$\%$     \\
\hline
\hline
Barred in groups      &  319    &   38.2$\%$     \\
Unbarred in groups    &  519    &   61.8$\%$     \\
\hline
Total              & 838     &   100.0$\%$     \\
\hline
\end{tabular}
{\small}
\end{table}

\begin{table}
\center
\caption{Percentages of barred and unbarred active galaxies with different color and stellar mass ranges.}
\begin{tabular}{|c c c| }
\hline
Color ranges &  Barred AGN & Unbarred AGN  \\
\hline
\hline
0.55$<(M_g-M_r)<$0.7   &  26.6$\%$    &   73.4$\%$     \\
0.7$<(M_g-M_r)<$0.85  &  45.4$\%$      &   54.6$\%$     \\
0.85$<(M_g-M_r)<$1.0  &  42.5$\%$      &   57.5$\%$     \\
\hline
\hline
Stellar mass ranges &   &   \\
\hline
10.2$<log(M_*/M_{\sun})<$10.6  &  31.2$\%$    &   68.8$\%$     \\
10.6$<log(M_*/M_{\sun})<$11.0  &  40.0$\%$    &   60.0$\%$     \\
11.0$<log(M_*/M_{\sun})<$11.4  &  40.5$\%$    &   59.5$\%$     \\
\hline
\end{tabular}
{\small}
\end{table}

In addition, we calculated the bar fraction as a function of colors and stellar masses of the host galaxies in our sample (see Table 2). We can observe that the bar fraction increases towards redder and more massive AGN host galaxies,
in accordance with previous work \citep{master11,oh12}.

\subsection{Control sample}

Some studies of the effect of bars from observational analysis show contradictory conclusions.
In several cases, authors have attempted to isolate the effects of bars by comparing barred galaxies with unbarred ones. 
However, different authors have proposed different ways to build comparison samples, and so the discrepancy in the results could be mainly due to a biased selection of these control samples.
By using SDSS mock galaxy catalogs built from the Millennium Simulation, \cite{perez09} showed that a suitable control sample for pair galaxies should be selected that at least match, redshift, morphology, stellar masses, and local density environments. 
This criteria is also applicable to the building of control samples of barred galaxies with the purpose of analysing different properties with respect to unbarred ones.

\begin{figure}
\includegraphics[width=87mm,height=130mm ]{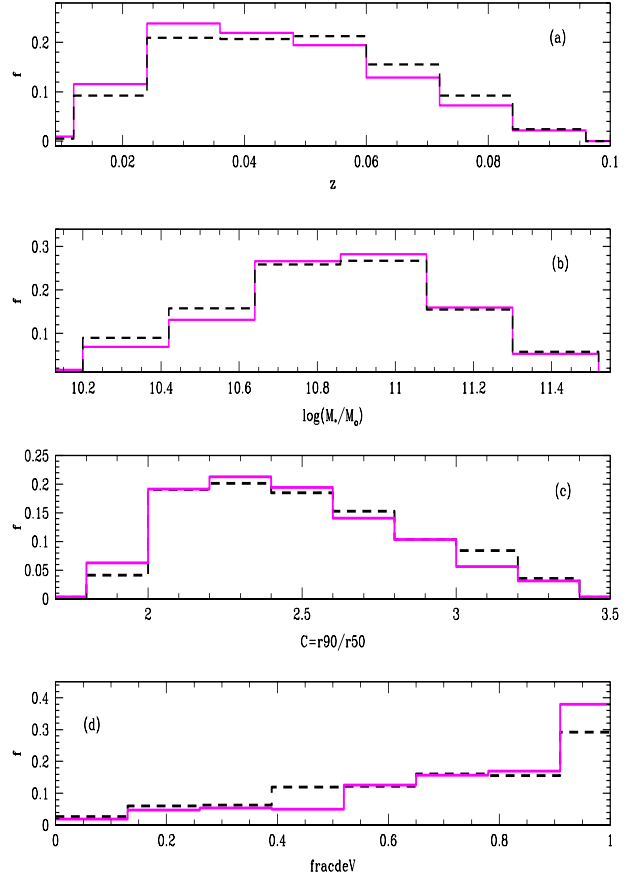}
\caption{Distributions of redshift ($z$), stellar mass ($log(M_*)/M_{\sun}$), concentration index ($C$),  
and bulge size indicator ($fracdeV$) (panels $a$, $b$, $c$, and $d$), for barred AGN host galaxies (solid lines) and active galaxies without bars in the control sample (dashed lines).}
\label{cont}
\end{figure}

Our aim is to focus on the effects of bars on the active nuclei galaxies that reside in high density environments. 
Therefore, we constructed a reliable control sample of unbarred AGN host galaxies that share similar environmental conditions in order to compare these with the barred host results.
Using a Monte Carlo algorithm we selected unbarred active galaxies in groups 
with similar redshift and stellar mass distributions 
(see panels $a$ and $b$ in Fig.~\ref{cont}). 
We have also considered unbarred AGNs in the control sample with similar concentration indices\footnote{$C=r90/r50$ is the ratio of Petrosian 90 \%- 50\% r-band light radii}, $C$, as for the barred AGN host sample, with the purpose of obtaining a similar bulge-to-disk ratio (see panel $c$). 
Thus, the differences in the results will be driven by the presence of bars and not by the difference in the morphology of the galaxies.
In addition, we have take into account active galaxies with similar bulge prominences in both AGN samples, selecting control galaxies
with similar distributions of the $fracdeV$ parameter to the
barred AGN host sample (panel $d$ in Fig.~\ref{cont}). 
This parameter is a good indicator of bulge sizes in disk galaxies 
\citep{ber10,kuehn05,master11,skibba12}.
The methodology followed to obtain the control sample of unbarred active galaxies in this work ensures that it will have the same selection effects as the barred AGN host catalog. It can then be used to estimate
the real difference between active galaxies with and without
bars in groups and clusters, unveiling the effect of bars on the nuclear activity and the role played by high density environments.

\section{Analysis of the relative position in groups}

\begin{figure}
\includegraphics[width=87mm,height=87mm ]{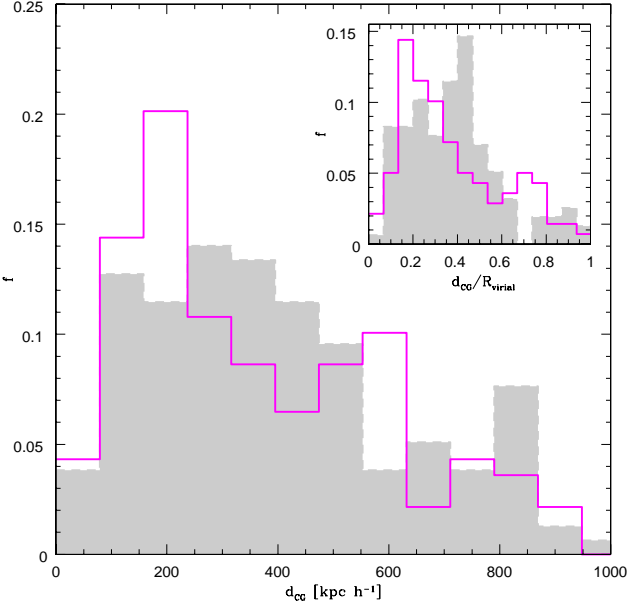}
\caption{Distributions of group-centric galaxy distances, $d_{CG}$, for
barred AGN (solid lines) and unbarred AGN (full surface) in the control sample.
The inset shows to the distribution of the normalized group-centric distance, $d_{CG}/R_{Virial}$, 
in the same samples.
}
\label{histdc}
\end{figure}

The study of the bar fraction in the extremely dense regions of the cluster core with respect to the lower density environments of the outer regions of the galaxy clusters can provide important clues to understanding the relative effects of internal and external physical mechanism on the evolution of the barred disk galaxies. 
We may ask whether barred AGN galaxies have a particular location in groups and clusters with respect to the control sample.
Therefore, we analyze the distribution of projected radial distance with respect to group centers, $d_{CG}$, for barred AGN galaxies and AGN without bars in the control sample.
In order to calculate reliable group centers, we have only considered groups with a minimum number of seven members, 
thereby avoiding large uncertainties in this determination. 
Both the virial masses and virial radius distributions of the restricted sample are similar to the total group sample.
The resulting distributions are shown in Fig.~\ref{histdc}.
It can be clearly seen that barred AGN galaxies are more concentrated towards the group center 
than the unbarred AGN in the control sample.
We also calculated the normalized group-centric distance, $d_{CG}/R_{Virial}$, finding the same 
tendencies (see the inset).
In addition, we find that the fraction of barred AGN galaxies nearest the group-centric distance,  $d_{CG}<300kpc$ $h^{-1}$, is higher ($\approx$ 49$\%$) with respect to those in the unbarred AGN sample ($\approx$ 38$\%$).

The pioneer work of \cite{thom81} also showed that the bar fraction is significantly larger in the core of the Coma cluster than in the outer region.

More recently, \cite{marinova12}, using high resolution images from the Hubble Space Telescope ACS Treasury survey of the Coma cluster at $z \approx 0.02$, studied bars in the dense core of this cluster.
They found a hint of an increase in the mean bar fraction toward the core of the Coma.
In addition, the study of the bars in lenticular galaxies (S0) in the rich Coma cluster showed that there is a radial increase in the bar fraction towards the cluster core with respect to the low density outer regions \citep{lans14}.  
\cite{and96} found that the radial distribution of the barred spirals in the Virgo cluster is clearly different than that of the spirals without bars, showing that barred galaxies are more centrally condensed than the unbarred ones.
Moreover, observational studies in several galaxy clusters also found an increase in the bar fraction in the dense core of the clusters \citep{and96,bar09,marinova12}.  

From numerical studies, \cite{byrd90} argued that the observed tendency for disk galaxies in the center of rich clusters to be barred is due to the strong tidal field. This can transform unbarred spiral galaxies into barred ones.
In this context, the tidal field has a constant direction and is symmetric about the spiral galaxy, as would be the case for cluster members approaching in an eccentric orbit towards the cluster center. 
This symmetric perturbation can produce distortions in the galaxy disk and then form strong arms and bars. 
In addition, other tidal effects of the cluster field have been studied, e.g., tidal stripping \citep{merritt84} and tidal disruption \citep{miller86}. These authors also found that galaxies in the core of the groups and clusters can survive over the long term without being destroyed. 
Our results confirm these previous findings for statistically reliable observational data.

\section{Color in host AGN galaxies}

Galaxy colors, as is well known, are correlated with star formation, age, gas content, and environment, all of which may be related to the presence of disk instabilities such as bars, and with the central nuclear activity. 
With the aim of exploring the colors of the barred active nuclei galaxies and the effects of denser regions, in this section we analyze colors in both samples of AGN with and without bars that inhabit groups and clusters. 

\begin{figure}
\includegraphics[width=80mm,height=90mm ]{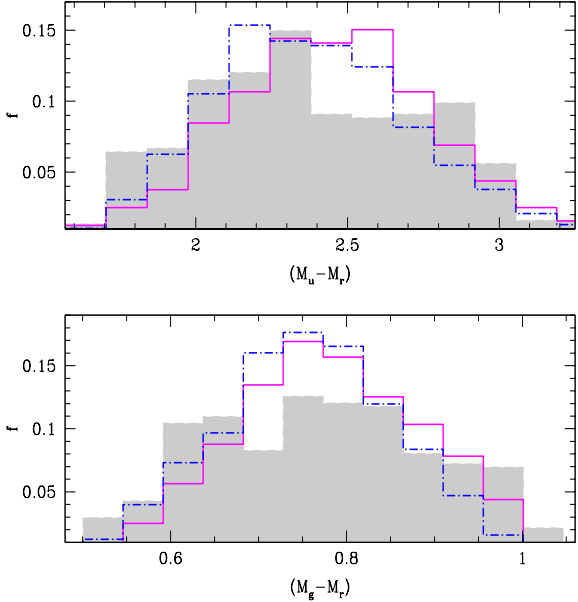}
\caption{Distribution of colors, $(Mu-Mr)$ and $(Mg-Mr)$ (upper and lower panels), 
for barred AGN host galaxies (solid lines) and AGN hosts without bars (shaded histograms)
within groups and clusters.
Dot-dashed lines represent barred AGN hosts in the field.
}
\label{histCol}
\end{figure}

In Fig.~\ref{histCol} we show the $(Mu-Mr)$ and $(Mg-Mr)$  color distributions of 
barred and unbarred active galaxies residing in groups and clusters. 
It can be clearly seen that there is a significant excess of barred AGN hosts 
with red colors. 
In Table 3 we quantify the excess of red color indexes ($(Mu-Mr)>2.4$ and $(Mg-Mr)>0.65$) 
of barred AGN hosts with respect to the unbarred AGN in the control sample.

Several previous studies displayed an excess of barred galaxies with redder colors from different samples.
\cite{master11} found a high fraction of bars in the passive red spiral galaxies for a sample obtained from the Galaxy Zoo catalog.
In addition, \cite{oh12} showed that a significant number of barred
galaxies are redder than their counterparts of unbarred spiral galaxies.
Furthermore, in our previous work (A13) we found an excess of red colors 
in a sample of isolated spiral barred AGN with respect to unbarred active galaxies in a suitable control sample.
The results obtained here from barred AGN hosts in high density environments are consistent with those of barred galaxies in low density regions.

In addition, we analyze the colors of barred AGN hosts in denser regions in comparison with those of active galaxies with bars in the field. For this purpose we 
used isolated barred AGN hosts obtained from our previous work (see A13 for details). This sample has similar redshift, stellar mass, concentration index, and bulge size indicator distributions to the sample of barred active galaxies in high density environments.
From Fig.~\ref{histCol} we can see that there is an excess of barred AGN hosts with redder colors in groups and clusters with respect to isolated barred active galaxies. 
Furthermore, we calculated the excess of red color indexes for the isolated sample, $(Mu-Mr)>2.4$ and $(Mg-Mr)>0.65$, finding 46.5$\%$ and 84.8$\%$, respectively.
This result also shows that isolated barred AGN hosts are redder than unbarred active galaxies in groups, suggesting that bars 
determine changes in the host colors.

\begin{figure}
\includegraphics[width=95mm,height=90mm ]{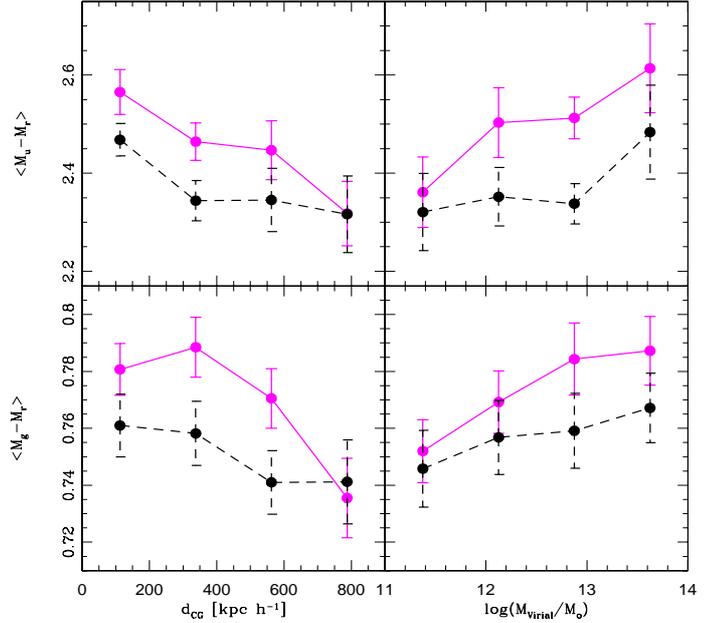}
\caption{Mean $<Mu-Mr>$ and $<Mg-Mr>$ galaxy colors (upper and lower panels) 
as a function of the group-centric galaxy distances, $d_{CG}$ (left panels), 
and virial group mass, $log(M_{Virial}/M_{\sun})$ (right panels), for barred and 
unbarred AGN host galaxies (solid and dashed lines).
}
\label{meancol}
\end{figure}

Moreover, to help us to understand the behavior of the colors in barred AGN hosts within high density environments, we measured the mean color values with respect to the group-centric distance.
In the left panels of Fig.~\ref{meancol}, we show the mean $<Mu-Mr>$ (upper panel) and $<Mg-Mr>$ (lower panel) colors as a function of group-centric distance for both samples 
of barred and unbarred active galaxies.
As is well known, the central regions of the groups and clusters of galaxies are mainly populated by objects redder than those in the outer regions.
It can also be seen that the number of active galaxies with redder colors increases toward the group centers, and this tendency is more significant in barred AGN with respect to those in the control active galaxies without bars.

Figure~\ref{meancol} (right panels) shows the mean colors for barred AGN (solid lines) as a 
function of the host group virial mass. 
We also show the results for the control sample (dashed lines).
As can be seen, the number of redder AGN hosts increases towards higher group virial masses.
In addition, we can see that barred AGN have systematically redder colors than active galaxies without bars in the control sample.

Redder colors found in barred active galaxies, with respect to unbarred counterparts,   
suggest that bar perturbations have a considerable effect on colors of AGN hosts.

\begin{table}
\center
\caption{Percentages of barred and unbarred AGN host galaxies in groups with red colors.}
\begin{tabular}{|c c c | }
\hline
Restrictions & $\%$ of barred &  $\%$ of unbarred \\
\hline
\hline
$(Mu-Mr)>2.4$   &   56.2$\%$ & 40.3$\%$  \\
$(Mg-Mr)>0.65$  &   89.6$\%$ & 75.7$\%$  \\
\hline
\end{tabular}
{\small}
\end{table}

\section{Local environmental properties}

\begin{figure}
\includegraphics[width=75mm,height=82mm ]{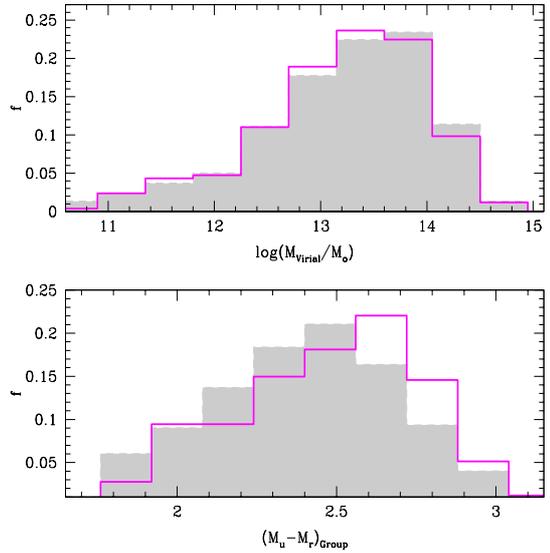}
\caption{Distributions of group virial masses, $log(M_{Virial}/M_{\sun})$ (upper panel), 
and group colors, $(M_u-M_r)_{Group}$ (lower panel), for host groups of barred and unbarred AGN galaxies  
(solid lines and full surface, respectively).
}
\label{Mvcol}
\end{figure}

\begin{figure}
\includegraphics[width=85mm,height=140mm ]{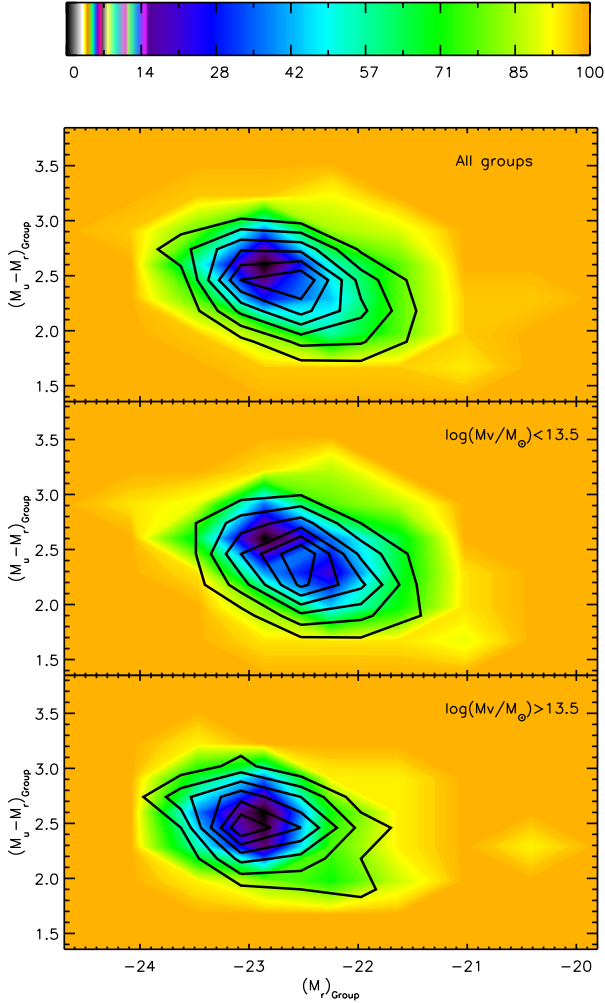}
\caption{Color-magnitude diagrams $(M_u-M_r)_{Group}$ versus $(M_r)_{Group}$. 
The density map shows the host groups of barred AGN (color scale corresponding to the cumulative percentage of objects, as shown in the key).  
For comparison, the black solid lines enclose 14$\%$, 28$\%$, 42$\%$, 57$\%$, 71$\%$, and 85$\%$ of the host groups of unbarred AGN in the control sample.
Middle and lower panels show host groups with different ranges of virial masses: 
$M_{Virial}<10^{13.5} M_{\sun}$ and $M_{Virial}>10^{13.5} M_{\sun}$, respectively.
}
\label{colmr}
\end{figure}

In this section, we analysed the host group properties.
For this purpose we explore the virial masses and the group colors in both host groups
of barred and unbarred active galaxies. 
We calculated the group color,  
$(M_u-M_r)_{Group}$, which is measured by summing
the luminosities of the four brightest galaxy members in the $r$ and $u$ bands in each group. 
Different authors \citep{Eke04,pad04} found that 
this measure is a good parameter with which to quantify the global group luminosity.

The results are plotted in Fig.~\ref{Mvcol} where we show the distributions of the virial masses, $log(M_{Virial}/M_{\sun})$ (upper panel), and the group colors, $(M_u-M_r)_{Group}$ (lower panel), of the groups hosting, for barred AGN and their unbarred counterparts. 
We find that both host groups of the barred and unbarred AGN show similar virial mass distributions.
Interestingly, the group color distributions of both samples show significant differences. The host groups of the corresponding barred active galaxies show a larger fraction of red colors with respect to the host groups of the unbarred galaxies in the control sample.
We calculate the fraction of active galaxies with and without bars that reside in red groups, defined as $(M_u-M_r)_{Group}>2.4$ and find a percentage of about 62.2 $\%$  and 51.8 $\%$, respectively.

Figure~\ref{colmr} shows color-magnitude diagrams for host groups of barred active galaxies and unbarred AGN in the control sample.
In addition, the analysis was performed for two ranges of group virial masses: $M_{Virial}<10^{13.5} M_{\sun}$ and $M_{Virial}>10^{13.5} M_{\sun}$ (middle and lower panels, respectively).
As can be observed, for the same $(M_r)_{Group}$, the host group colors of barred active galaxies show  distributions spreading toward red populations with respect to the host group colors of the unbarred AGN objects. 
In addition, this trend is more significant in less massive groups.
However, groups with $M_{Virial}>10^{13.5} M_{\sun}$ have a more similar color-magnitude distributions than that of host groups of active galaxies without bars.


\section{Nuclear activity in barred AGN galaxies} 

With the aim of assessing the effect of bars on the nuclear activity and the role played by high density environments, 
in this section we analyzed the central activity in the samples of AGN galaxies with and without bars in groups and clusters.
We focused on the dust-corrected luminosity of the [OIII]$\lambda$5007 line, $Lum[OIII]$, as a tracer of the AGN activity.
The [OIII] line is the strongest narrow emission 
line in optically obscured AGN and has a low contamination from contributions of star formation lines in the host galaxy. 
Furthermore, the $Lum[OIII]$ estimator has been widely used by different authors in several works \citep{mul94, kauff03, heck04, heck05, brinch04}.
From an inspection of the BPT diagrams, \cite{kauff03} found that high metallicity hosts show a low contamination from star formation in the $Lum[OIII]$. 
In our sample, most of the galaxies have stellar masses $M^* >10^{10}$ $M_{\sun}$ 
(see panel $b$ in Fig.2), therefore their metallicities are expected to be high because of the mass-metallicity relation \citep{kauff03, tremonti04}, with low contamination of star formation lines. 

\begin{figure}
\includegraphics[width=70mm,height=60mm ]{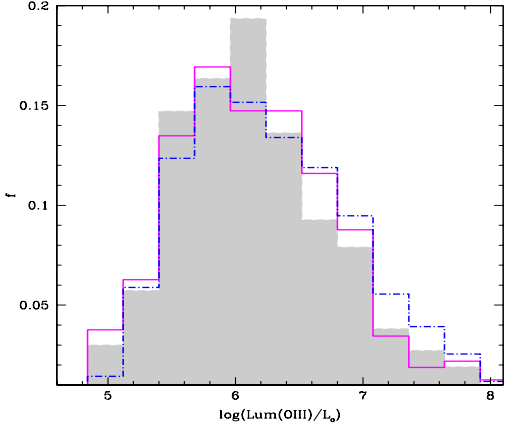}
\caption{Distributions of $log(Lum[OIII]/L_{\sun})$ for barred AGN (solid line) and unbarred active galaxies (full surfaces) in high density environments.
Dot-dashed line represents isolated barred AGN galaxies.}
\label{histLOF}
\end{figure}

In Fig.~\ref{histLOF} we show the distributions of $Lum[OIII]$ for AGN with and without bars within 
groups and clusters. 
We also plot isolated barred active galaxies residing in the field. 
It is interesting that in high density regions barred AGN show a slight trend toward higher values of $Lum[OIII]$ than AGN without bars in the control sample.
Moreover, isolated active galaxies with bars show an excess of higher nuclear activity values with respect to those of barred AGN within groups and clusters. 
Following A13, we divide the sample by using the  $Lum[OIII] $ to study trends with luminosity. Then,  we consider the luminosity $Lum[OIII] = 10^{6.4} L_{\odot}$ as a limit between the lower-luminosity and the higher-luminosity subsamples. Thus, we estimate that the $43.4\%$, $35.7\%$, and $31.5\%$ of isolated barred active objects, barred, and unbarred AGN hosts in denser regions, respectively, belong to the higher-luminosity subsample, because they are the strongest AGN.
These findings indicate that the central nuclear activity in barred AGN galaxies is more efficient in lower density regions than in higher density environments.

\begin{figure}
\includegraphics[width=100mm,height=115mm ]{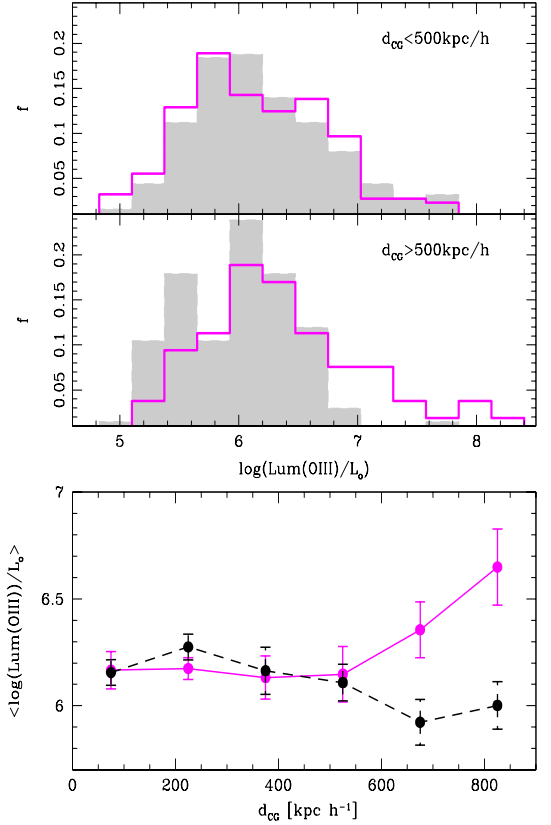}
\caption{Distributions of $log(Lum[OIII]/L_{\sun})$ for barred AGN (solid lines) and unbarred active galaxies (full surfaces) with different group-centric distances: 
$d_{CG}<500kpc$ $h^{-1}$ and $d_{CG}>500kpc$ $h^{-1}$ (upper and middle panels, respectively).
Lower panel shows mean $<log(Lum(OIII)/L_{\sun})>$ as a function of group-centric distance, $d_{CG}$, 
for AGN galaxies with and without bars in groups (solid and dashed lines, respectively).
}
\label{histLOIII}
\end{figure}

Figure~\ref{histLOIII} (upper and middle panels)  shows the nuclear activity distributions, $log(Lum[OIII]/L_{\sun})$, of barred active galaxies compared with the unbarred AGN host counterparts. 
The analysis was performed for the different ranges of group-centric distances 
($d_{CG}<500kpc$ $h^{-1}$ and $d_{CG}>500kpc$ $h^{-1}$),  
 with the purpose of studying the efficiency of nuclear activity in AGN galaxies with and without bars located in the cluster cores and in the regions far from the group centers.
Different authors \citep{sar80, ken82} found that the values of the cluster core radius are typically between 300 $kpc$ and 600 $kpc$. \cite{dres80} also obtained similar values for the sample of 55 rich clusters. Moreover, 
the X-ray emitting gas gives a suitable proxy to the mass distribution in the clusters, which is an important parameter for obtaining the core radius. 
In this context, several X-ray observations \citep{for82, pat00, haa10} usually imply core radii of $\approx$ 500 $kpc$.
We found that towards the group center, the $Lum[OIII]$ distributions of barred AGN hosts are similar to those of AGN without bars in the control sample.
Interestingly, towards regions farthest from the group center,
the activity distributions of barred and unbarred AGN hosts exhibit significant differences, with barred objects having a higher fraction of powerful AGN.
We have also computed the mean $<log(Lum(OIII)/L_{\sun})>$ values for different bins of group-centric distance, $d_{GC}$, 
for both AGN samples.
The results from the lower panel in Fig.~\ref{histLOIII} show clearly
that, in barred AGN for larger group-centric distance, the $Lum[OIII]$ values strongly increases, while in active galaxies without bars in the control sample remains approximately constant with a slight decline.
It could be interpreted in terms of the greater ability of bars to fuel the central black holes in host group member galaxies located far from the group center.

\begin{figure}
\includegraphics[width=100mm,height=115mm ]{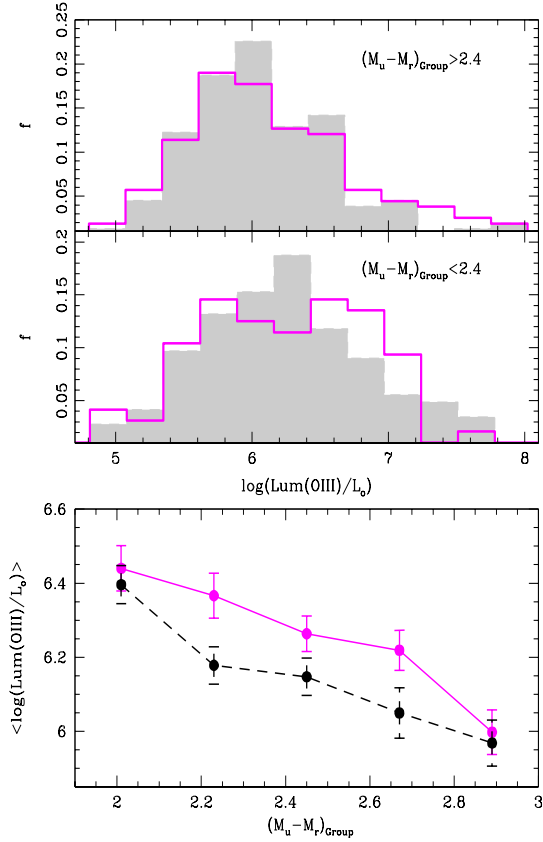}
\caption{Distributions of $log(Lum[OIII]/L_{\sun})$ for barred AGN (solid lines) and unbarred active galaxies (full surfaces) with different host group color ranges: 
$(M_u-M_r)_{Groups}>2.4$ and $(M_u-M_r)_{Groups}<2.4$ (upper and middle panels, respectively).
Lower panel shows mean $log(Lum(OIII)/L_{\sun})$ as a function of the host group colors for
 AGN galaxies with and without bars in groups (solid and dashed lines, respectively).
}
\label{LODC}
\end{figure}

With the aim of understanding the behavior of the nuclear activity in AGN galaxies with and without bars as a function of host group colors, in Fig.~\ref{LODC} we show the nuclear activity distributions for different ranges of group colors:  
$(M_u-M_r)_{Groups}>2.4$ and $(M_u-M_r)_{Groups}<2.4$ (upper and middle panels, respectively).
It can be clearly seen that barred AGN show a trend toward higher values of $Lum[OIII]$ than AGN without bars in the control sample. This tendency is more significant in active galaxies that inhabit bluer groups.
Moreover, in Table 4 we quantify the percentage of barred and unbarred active galaxies in denser regions with 
$Lum[OIII] = 10^{6.4} L_{\odot}$, for different ranges of group-centric distance and host group colors.

In addition, in Fig.~\ref{LODC} (lower panel) we present the mean 
$<log(Lum[OIII]/L_{\sun})>$ values as a function of the global group colors.
This relation was calculated for the corresponding host groups of the barred AGN and for active galaxies without bars in the control sample. 
The results clearly show that generally in bluer groups, active galaxies show higher $Lum[OIII]$ values.
We can also see that barred AGN objects systematically show higher nuclear activity, irrespective of the group colors. 
These findings show that the higher-luminosity subsample AGN reside preferentially in bluer groups, which could provide suitable conditions for the central black hole feeding.  
In this context, the most efficient nuclear activity in barred AGN galaxies with respect to their unbarred counterparts reflect that bars have an important role in helping the gas infall towards the central regions in active galaxies. 
Considering that the group colors, $(M_u-M_r)$, given by the four brightest members is a representative parameter of the 
global color of the groups, this result is in agreement with that found by several authors. 
\cite{cold09} show that the power of the AGN activity is strongly dependent on the environment, finding that the AGN with higher values of $<log(Lum[OIII]/L_{\sun})>$ are located in regions populated by bluer galaxies.
Similar results were found by \cite{Padilla10,cold14} suggesting that AGN very close to the high density regions need more available gas to effectively feed the central black hole. This is reflected in the higher probability of finding AGN in regions populated by bluer galaxies.

\begin{table}
\center
\caption{Percentages of barred and unbarred active galaxies with $Lum[OIII]> 10^{6.4} L_{\odot}$ 
in different ranges of group-centric distances and host group colors. 
}
\begin{tabular}{|c c c| }
\hline
Ranges & $\%$ of barred AGN & $\%$ of unbarred AGN \\
\hline
\hline
$d_{CG}<500kpc$ $h^{-1}$   & 34.1$\%$  &     34.4$\%$        \\
$d_{CG}>500kpc$ $h^{-1}$   & 41.5$\%$  &     20.8$\%$        \\
\hline
$(M_u-M_r)_{Groups}>2.4$   & 31.1$\%$  &     26.3$\%$        \\
$(M_u-M_r)_{Groups}<2.4$   & 43.8$\%$  &     34.2$\%$        \\
\hline
\end{tabular}
{\small}
\end{table} 

\section{Summary and conclusions}

We have performed a statistical analysis of local environmental properties, host characteristics, and nuclear activity of the AGN spiral galaxies with and without bars within denser regions.
For this purpose, we have identified barred and unbarred objects that inhabit groups and clusters of galaxies, by cross-correlating the total barred and unbarred AGN catalogs obtained from our previous work (see Alonso et al. 2013 for details) with the SDSS-DR7 group catalog constructed by \cite{zap09}.
To obtain an appropriate quantification of the effects of bars on the active nuclei and the role played by high density environments, we constructed a suitable control sample of unbarred AGN galaxies within groups and clusters, with the same redshift, stellar mass, concentration index, and bulge size parameter distributions. 

We now summarize the principal results of our analysis and its main conclusions.

{\em(i)} We found that the fraction of barred AGN galaxies in groups ($\approx$ 38$\%$) is higher than those in the total barred AGN sample ($\approx$ 28$\%$). 
This result shows that AGN spiral galaxies in groups are more likely to be barred than those in the field, 
indicating that galactic bars are stimulated in accordance with the local density environment, and also that the physical mechanisms produced within groups and clusters of galaxies (ram pressure, strangulation, harassment and  interactions) seem to be related to the bar phenomenon
\citep{kor12,bere07,master12}. 
In addition, we observed that the bar fraction increases towards redder and more massive AGN host galaxies.

{\em(ii)} We have found that barred AGN galaxies are more concentrated towards the group centers than the other unbarred AGN galaxy members. We quantified this trend and we found that nearest the group-centric distance, $d_{CG}<300kpc$ $h^{-1}$, the fraction of barred AGN galaxies is higher ($\approx$ 49$\%$) with respect to those in the unbarred AGN sample ($\approx$ 38$\%$).
This implies that barred phenomenon are more likely to occur in the dense core of the groups and clusters, where the  strong tidal field can transform unbarred spiral galaxies into barred ones (Byrd \& Valtonen 1990).

{\em(iii)} We examined the color distributions of barred and unbarred AGN host galaxies within groups and clusters, and we found that barred AGN hosts have a clear excess of redder colors than
galaxies in the control sample.
In addition, we analyzed the colors of AGN with bars in the field, finding that isolated barred AGN hosts are redder than active galaxies without bars in groups.
This result suggests that bar perturbations can significantly affect galaxy colors in the hosts of AGN.

{\em(iv)} We also studied the mean colors in barred AGN galaxies and in the control sample as a function of the group-centric distance and group virial masses.
We found that the number of red galaxies increases toward the group center as expected, and 
barred galaxies have a higher fraction of redder colors than active galaxies without bars within groups/clusters.
We also show that redder galaxies increases towards higher group virial masses and, in this context, barred AGN galaxies have  systematically redder colors than active galaxies without bars in the control sample, for all virial mass ranges.

{\em(v)} We explored the host group properties of the AGN galaxies with and without bars, and we found that both host groups show similar virial masses. However, the group color distributions of both samples present significant differences; namely, the host groups of the barred AGN exhibit a larger fraction of red colors, with respect to the host groups of the corresponding unbarred active galaxies in the control sample.
We also examined the color-magnitude relation for both host groups and we found that at the same $(M_r)_{Group}$, 
the host group colors of barred active galaxies display distributions spreading toward red populations with respect to the host groups of the unbarred AGN objects. 
This trend is more significant in less massive groups than in 
groups with $M_{Virial}>10^{13.5} M_{\sun}$.

{\em(vi)} We found that in high density regions, barred active galaxies show an excess of nuclear activity compared to galaxies without bars in the control sample.
 We have also analyzed barred active galaxies in the field, and we found that isolated AGN galaxies with bars 
show more efficient nuclear activity with respect to those of barred AGN within groups and clusters.

{\em(vii)} We also analyzed the relation between the central nuclear activity with the group-centric distances, and with the global group colors for barred active galaxies and unbarred AGN in the control sample. 
From this study, we conclude that, in particular,  
the nuclear activity has strongly increased in barred active galaxies located far of the group center, while the nuclear activity in AGN galaxies without bars remains approximately constant with the group-centric distance.
In addition, we found that in bluer host groups active galaxies show a higher $Lum[OIII]$ values for both AGN samples; however, barred active objects systematically show a higher nuclear activity, irrespective of the global group colors.
These results suggest that the efficiency of bars in transporting material towards the more central regions of the AGN galaxies in high density environments has a significant dependence on the localization of objects within the group and/or cluster and on the host group colors.

\begin{acknowledgements}
      This work was partially supported by the Consejo Nacional de Investigaciones
Cient\'{\i}ficas y T\'ecnicas and the Secretar\'{\i}a de Ciencia y T\'ecnica 
de la Universidad Nacional de San Juan.

Funding for the SDSS has been provided by the Alfred P. Sloan
Foundation, the Participating Institutions, the National Science Foundation,
the U.S. Department of Energy, the National Aeronautics and Space
Administration, the Japanese Monbukagakusho, the Max Planck Society, and the
Higher Education Funding Council for England. The SDSS Web Site is
http://www.sdss.org/.

The SDSS is managed by the Astrophysical Research Consortium for the
Participating Institutions. The participating institutions are the American
Museum of Natural History, Astrophysical Institute Potsdam, University of
Basel, University of Cambridge, Case Western Reserve University,
University of
Chicago, Drexel University, Fermilab, the Institute for Advanced Study, the
Japan Participation Group, Johns Hopkins University, the Joint Institute for
Nuclear Astrophysics, the Kavli Institute for Particle Astrophysics and
Cosmology, the Korean Scientist Group, the Chinese Academy of Sciences
(LAMOST), Los Alamos National Laboratory, the Max-Planck-Institute for
Astronomy (MPIA), the Max-Planck-Institute for Astrophysics (MPA), New Mexico
State University, Ohio State University, University of Pittsburgh, University
of Portsmouth, Princeton University, the United States Naval Observatory, and
the University of Washington.
\end{acknowledgements}

\end{document}